\documentclass[sigconf]{acmart}
\AtBeginDocument{%
  \providecommand\BibTeX{{%
    \normalfont B\kern-0.5em{\scshape i\kern-0.25em b}\kern-0.8em\TeX}}}

\setcopyright{acmcopyright}
\copyrightyear{2022}
\acmYear{2022}
\acmDOI{XXXXXXX.XXXXXXX}

\acmConference[CIKM '22]{CIKM}{October 17--22, 2022}{Atlanta, GA}
\acmPrice{15.00}
\acmISBN{978-1-4503-XXXX-X/18/06}

\begin{document}

\title{EvalRS: a Rounded Evaluation of Recommender Systems}

\author{Jacopo Tagliabue}
\affiliation{%
  \institution{Coveo Labs}
  \city{New York}
  \country{USA}
}
\authornote{\textbf{Contributions}: TS proposed the format and methodology and worked with JT and FB towards a first draft. PC led the implementation and contributed most of the \texttt{RecList} code. GA, CG, FB and PC researched, iterated and operationalized behavioral tests. GM reviewed the API and implemented baselines, while GA, JT and FB prepared tutorials for participants. \textit{Everybody} helped with drafting the paper, rules and guidelines. JT and FB acted as senior PIs in the project.}

\author{Federico Bianchi}
\affiliation{%
  \institution{Stanford University}
  \city{Stanford}
  \country{USA}}

\author{Tobias Schnabel}
\affiliation{%
  \institution{Microsoft}
  \city{Redmond}
  \country{USA}}
  
\author{Giuseppe Attanasio}
\affiliation{%
  \institution{Bocconi University}
  \city{Milan}
  \country{Italy}}
  
\author{Ciro Greco}
\affiliation{%
  \institution{Coveo Labs}
  \city{New York}
  \country{USA}
}

\author{Gabriel de Souza P. Moreira}
\affiliation{%
  \institution{NVIDIA}
  \country{Brazil}}
  
\author{Patrick John Chia}
\affiliation{%
  \institution{Coveo}
  \city{Montreal}
  \country{Canada}}  

\renewcommand{\shortauthors}{Tagliabue et al.}

\begin{abstract}
 Much of the complexity of Recommender Systems (RSs) comes from the fact that they are used as part of more complex applications and affect user experience through a varied range of user interfaces. However, research has focused almost exclusively on the ability of RSs to produce accurate item rankings while giving little attention to the evaluation of RS behavior in real-world scenarios. Such narrow focus has limited the capacity of RSs to have a lasting impact in the real world and makes them vulnerable to undesired behavior, such as reinforcing data biases. We propose \texttt{EvalRS} as a new type of challenge, in order to foster this discussion among practitioners and build in the open new methodologies for testing RSs ``in the wild''.
\end{abstract}



\maketitle

\section{Introduction}

Recommender systems (RSs) are embedded in most applications we use today. From streaming services to online retailers, the accuracy of a RS is a key factor in the success of many products. Evaluation of RSs has often been done considering point-wise metrics, such as \textit{Hit Rate} or \textit{nDCG} over held-out data points, but the field has recently begun to recognize the importance of a more rounded evaluation as a better proxy to real-world performance \cite{DBLP:journals/corr/abs-2111-09963}. 

We designed \texttt{EvalRS} as a new type of data challenge, in which participants are asked to test their models incorporating quantitative as well as behavioral insights: starting from a popular open dataset -- \textit{Last.fm} --, we devise evaluation metrics that are meant to test properties of recommender systems that go beyond aggregate numbers on a test set. The contribution of this challenge is therefore two-fold:

\begin{enumerate}
    \item we propose and standardize the data, evaluation loop and testing for RSs over a popular use case (user-item recommendations for music consumption \cite{10.1145/2911996.2912004}), thus releasing in the open domain a first unified benchmark for this topic;
    \item we bring together the community on evaluation from both an industrial and research point of view, to foster an inclusive debate for a more nuanced evaluation of RSs.
\end{enumerate}

In \textit{this} paper, we describe the philosophical and practical motivations behind \texttt{EvalRS}, provide context on the organizers, related events and relevant literature, and explain the evaluation methodology we champion. For participation rules, up-to-date implementation details and all the artifacts produced before and during the challenge, please refer to the \texttt{EvalRS} official repository.\footnote{ \url{https://github.com/RecList/evalRS-CIKM-2022}.}

\section{Motivation}
\label{sec:motivation}

\texttt{EvalRS} at \textit{CIKM 2022} complements the existing challenge landscape and it is driven by two different perspectives: the first one coming from academic research, the second one from the industrial development of RSs. We examined these in turn.

\subsection{A Research Perspective}

Although undeniable progress was made in the past years, concerns have been raised about the status of research advancements in the field of recommendations, particularly with respect to ephemeral processes in motivating architectural choices and lack of reproducibility ~\cite{10.1145/3298689.3347058}. \textit{This} challenge draws attention to a further  -- and potentially deeper -- issue: even if the ``reproducibility crisis'' is solved, we are still mostly dealing with  point-wise quantitative metrics as the only benchmarks for RSs. As reported by \citet{sun2020we}, the dominating metrics used in the evaluation of recommender systems published at top-tier conferences (RecSys, SIGIR, CIKM) are typical Information Retrieval metrics, such as \textit{MRR}, \textit{Recall}, \textit{HITS}, \textit{NDCG}~\cite{wang2019neural,Rashed2020MultiRecAM,10.1145/3383313.3412235,10.1145/3383313.3412263,bianchi-etal-2021-query2prod2vec}. 

While it is undoubtedly convenient to summarize the performance of different models via \textit{one} score, this is a ``lossy compression'' that discards a lot of important information on model behavior: for example, given the power-law distribution in many real-world datasets (\cite{CoveoSIGIR2021,10.1145/2827872,10.1145/3344257}), marginal improvements on frequent items may translate in noticeable accuracy gains, even at the cost of significantly degrading the experience of subgroups. Metrics such as coverage, serendipity, and bias \cite{Kotkov2016ChallengesOS,jannach2017recurrent,ludewig2018evaluation} are a first step in the right direction, but they still fall short of capturing the full complexity of deploying RSs.

Following the pioneering work of \cite{Ribeiro2020BeyondAB} in Natural Language Processing, we propose to supplement standard retrieval metrics with new tests: in particular, we encourage practitioners to go beyond the false dichotomy ``quantitative-and-automated'' vs ``qualitative-and-manual'', and find a middle ground in which behavioral \textit{desiderata} can be expressed transparently in code~\cite{DBLP:journals/corr/abs-2111-09963}.

\subsection{An Industrial Perspective}
RSs in practice differ from RSs used in research in crucial ways. For example, in research, a static dataset is used repeatedly, and there is no real interactivity between the model and users: prediction over a given point in time $x_t$ in the test set doesn't change what happens at $x_{t+1}$\footnote{This is especially important in the context of sequential recommender\cite{transformers4rec2021}, which arguably resembles more reinforcement learning than supervised inference with pseudo-feedback \cite{Ariu2020RegretIO}.}. Even without considering the complexity of reproducing real-world interactions for benchmarking purposes, we highlight four important themes from our experience in building RSs at scale in production scenarios:

\begin{itemize}
    \item \textit{Cold-start performance}: new/rare items and users are challenging for many models across industries~\cite{10.1145/3383313.3411477, 2106.03819}. In e-commerce, for instance, while most ``similar products'' predictions will happen over frequent items, in reality, new users and items can represent a big portion of them with significant business consequences: the cold-start problem is believed to affect ~50\% of users \cite{Hendriksenetal2020} in a context where field studies found that ~40\% of shoppers would stop shopping if shown non-relevant recommendations \cite{emarketer}.
    \item \textit{Use cases and industry idiosyncrasies}: different use cases in different industries present different challenges. For instance, recommendations for complementary items in e-commerce need to account for the fact that if item A is a good complementary candidate for item B, the reverse might not hold (e.g. an \textit{HDMI cable} is a good complementary item for a \textit{4k TV}, but not vice versa).
    Music recommendations need to deal with the issue of ``hubness'', where popular items act as hubs in the top-N recommendation list of many users without being similar to the users' profiles and making other items invisible to the recommender~\cite{Flexeretal2012}. Such use-case specific traits are particularly important when designing effective testing procedures and often require considerable domain knowledge. 
    \item \textit{Not all mistakes are equal}: point-wise metrics are unable to distinguish different \textit{types} of mistakes; this is especially problematic for recommender systems, as even a single mistake may cause great social and reputational damage \cite{NYTIMES_AMAZON}.
    \item \textit{Robustness matters as much as accuracy}: while historically a significant part of industry effort can be traced back to a few key players, there is a blooming market of Recommendation-as-a-Service systems designed to address the needs of ``reasonable scale'' systems~\cite{10.1145/3460231.3474604}. Instead of vertical scaling and extreme optimization, SaaS providers emphasize horizontal scaling through multiple deployments, highlighting the importance of models that prove to be flexible and robust across many dimensions (e.g., traffic, industry, etc.).
\end{itemize}

While not related to model evaluation \textit{per se}, decision-making processes in the real world would also take into account the different resources used by competing approaches: time (both as time for training and latency for serving), computing (CPU vs GPU), CO2 emissions are all typically included in an industry benchmark.

\section{EvalRS Challenge}
\label{sec:challenge}

\subsection{Improving testing for RSs}
We propose to supplement standard retrieval metrics over held out data points with \textit{behavioral tests}: in behavioral tests, we treat the target model as a black-box and supply only input-output pairs (for example, query user and desired recommended song). In particular, we leverage a recent open-source package, \texttt{RecList} \cite{DBLP:journals/corr/abs-2111-09963}, to prepare a suite of tests for our target dataset (Section \ref{sec:dataset}). In putting forward our tests, we operationalize the intuitions from Section~\ref{sec:motivation} through a general plug-and-play API to facilitate model comparison and data preparation, and by providing convenient abstractions and ready-made recommenders used as baselines. 

\subsection{Use Case and Dataset}

\texttt{EvalRS} is a user-item recommendation challenge in the music domain: participants are asked to train a model that, given a user id, recommends an appropriate song out of a known set of songs. The ground truth necessary to compute all the test metrics, quantitative \textit{and} behavioral, is provided by our leave-one-out framework: for each user, we remove a song from their listening history and use it as the ground truth when evaluating the models.

\label{sec:dataset}
We provide test abstractions and an evaluation script designed for \textit{LFM}, a transformed version of \textit{LFM-1b dataset} \cite{10.1145/2911996.2912004} -- a dataset focused on music consumption on \textit{Last.fm}. We chose the \textit{LFM-1b dataset} as the primary data source after a thorough comparisons of popular datasets for a unique combination of features: for example, given our focus on rounded evaluation and the importance of joining prediction / ground truth with meta-data, \textit{LFM} is an ideal dataset, as it provides rich song (artist, album information) and user (country, age, gender,\footnote{Gender in the original dataset is a binary variable. This is a limitation, as it gives a stereotyped representation of gender. Our intent is not to make normative claims about gender.} time on platform) meta-data.


We applied principled data transformations to make \texttt{EvalRS} amenable to a larger audience whilst preserving the rich information in the original dataset. We detail the data transformation process and our motivations:

\begin{itemize}
    \item First, we removed \texttt{users} and \texttt{artists} which have few interaction since they are likely to be too sparse to be informative. Following the suggestions in, we apply k-core~\cite{batagelj2002generalized} filtering  to the bipartite interaction graph between \texttt{users} and \texttt{artists}, setting $k=10$ (i.e. we retain vertices with a minimum degree of k).
    \item After the aforementioned processing, the dataset still contained over 900M events, which motivated further filtering of the data. In particular, we keep only the \textit{first} interaction a \texttt{user} had with a given \texttt{track}, and for each \texttt{user} we retain only their $N=500$ most recent unique \texttt{track} interactions. We supplement the information lost during this pruning step by providing the interaction count between a \texttt{user} and a \texttt{track}.
    \item We then performed another iteration of k-core filtering, this time on the \texttt{user-track} interaction graph, with $k=10$ to retain only users and tracks which are informative.
    \item Lastly, the original dataset contained missing meta-data (e.g. there were \texttt{track\_id} in the events data which did not have corresponding track metadata). We removed tracks, albums, artists and events which had missing information.
    \item We summarize the final dataset statistics in Table \ref{table:dataset_stats}.
\end{itemize}

\begin{table}
\centering
\begin{tabular}{ |l|c| } 
 \hline
 Items & Value \\
 \hline
 Users & $119,555$ \\ 
 Artists & $62,943$ \\ 
 Albums & $1,374,121$ \\ 
 Tracks & $820,998$ \\ 
 Listening Events & $37,926,429$\\ 
 User-Track History Length (25/50/75 pct) & $241/346/413$ \\ 
 \hline
\end{tabular}
\label{table:dataset_stats}
\caption{\label{table:dataset_stats} Descriptive statistics for \textit{LFM} dataset.}
\vspace{-8mm}
\end{table}

Taken together, these features allow us to fulfill \texttt{EvalRS} promise of offering a challenging setting and a rounded evaluation. While a clear motivation behind the release of \textit{LFM-1b dataset} was to offer ``additional user descriptors that reflect their music taste and consumption behavior'', it is telling that both the modelling and the evaluation by the original authors are still performed without any real use of these rich meta-data \cite{Schedl2017InvestigatingCM}. By taking a fresh look into an existing, popular dataset, \texttt{EvalRS} challenges practitioners to think about models not just along familiar quantitative dimensions, but also along non-standard scores closer to human perception of relevance and fairness. 

\subsection{Evaluation Metrics}
\label{sec:eval}

Submission are evaluated according to our randomized loop (Section \ref{sec:methdology}) over the testing suite released with the challenge. At a first glance, tests can be roughly divided in three main groups:

\begin{itemize}
    \item \textbf{Standard RSs metrics}: these are the typical point-wise metrics used in the field (e.g. MRR, HR@K) -- they are included as sanity checks and as a informative baseline against which insights gained through the behavioral tests can be interpreted. 
    \item \textbf{Standard metrics on a per-group or slice basis}: as shown for example in \cite{DBLP:journals/corr/abs-2111-09963}, models which are indistinguishable on the full test set may exhibit very different behavior across data slices. It is therefore crucial to quantify model performance for specific input and target groups, i.e. is there a performance difference between males and females? Is there an accuracy drops when artists are not very popular?
    \item \textbf{Behavioral tests}: this group may include perturbance tests (i.e. if we modify a user's history by swapping \textit{Metallica} with \textit{Pantera}, how much predictions will change?), and error distance tests (i.e. if the ground truth is \textit{Shine On You Crazy Diamond} and the prediction is \textit{Smoke on the Water}, ``how wrong'' is the model?).
\end{itemize}

Based on this taxonomy, we now survey the tests implemented in the \texttt{RecList} powering \texttt{EvalRS}, with reference to relevant literature and examples from the target datasets. For implementation details please refer to the official repository.\footnote{ \url{https://github.com/RecList/evalRS-CIKM-2022}.}

\subsubsection{Standard RSs metrics} Based on popular metrics in the literature, we picked two standard metrics as a quantitative baseline and sanity check for our \texttt{RecList}:

\begin{itemize}
    \item  \textbf{Mean Reciprocal Rank} (MRR) as a measure of where the first relevant element retrieved by the model is ranked in the output list. Besides being considered a standard rank-aware evaluation metric, we chose MRR because it is particularly simple to compute and to interpret.
    \item \textbf{Hit Rate} (HR), defined as \textit{Recall at k} ($k=100$), i.e. the proportion of relevant items found in the top-k recommendation.
\end{itemize}

\subsubsection{Standard metrics on a per-group or slice basis}
Models are tested to address a wide spectrum of known issues for recommender systems, for instance: fairness (e.g. a model should have equal outcomes for different groups, e.g. \cite{YangandStoyanovich2016, Castillo2019, Zehlikeetal2021}), robustness (e.g. a model should produce good outcomes also for long-tail items, such as items with less history or belonging to less represented categories, e.g. \cite{OMahony2018}), industry-specific use-cases (e.g. in the case of music, your model should not consistently penalize niche or simply less known artists).

All the tests in this group are based on \textit{Miss Rate} (MR), defined as ratio between the prediction errors (i.e. model predictions do not contain the ground truth) and the number of predictions. Slices can be generalized as\textit{n} partitions (e.g. Countries with UK/US/IT/FR and others is split is N partitions) of the test data forming \textit{n}-ary classes. The absolute difference between the MR obtained on each slice and the the MR obtained on the original test set is averaged and negated (so that a higher value implies better performance in the metric) to obtain the final score for each test. The slice-based tests considered for the final scores are:

\begin{itemize}
    \item \textbf{Gender balance}. This test is meant to address fairness towards gender \cite{SaxenaandJein2020}. Since the dataset only provides binary gender, the test will minimize the difference between the MR obtained on users who declared Female as gender and the MR obtained on the original test set. In other words, we operationalize this test as the smaller the difference: the fairer the model towards potential gender biases.
    \item \textbf{Artist popularity}. This test is meant to address a known problem in music recommendations: niche (or simply less known) artists and users who are less interested in highly popular content are often penalized by recommender systems \cite{Kowaldetal2020, CelmaandCano2008}. This point appears even more important when we consider that several music streaming services (e.g. Spotify, Tidal) also act as marketplaces for artists to promote their music.
    Since splitting the test set in two would draw an arbitrary line between popular vs. unpopular artists, failing to capture the actual properties of the distribution, the test set was divided with logarithmic bucketing in base 10.
    \item \textbf{User country}. Music consumption is subject to many country dependent factors, such as language differences, local sub-genres and styles, local licensing and distribution laws, cultural influences of local traditional music, etc. Since, as some argued, digitization has led to more diverse cultural markets \cite{BelloandGarcia2021}, these factors have deep implications for how people listen to music, how artists, labels and streaming platforms go to market. In this test, the test set was sliced selecting the top-10 countries based on the number of users.
    \item \textbf{Song popularity}. This test measures the model performance on both ``most popular'' tracks and on songs with fewer listening events. The test is designed to address both robustness to long tail items and cold-start scenarios, so we pooled together both less popular and newer songs. Also in this case, we used logarithmic bucketing in base 10 to divide the test set in order to avoid arbitrary thresholds.
    \item \textbf{User history}. The test can be viewed as a robustness/cold-start test, in which we sliced the dataset based on the length of user history on the platform:history is operationalized in terms of user play counts (i.e. the sum of play counts per user). Also in this case, we used logarithmic bucketing in base 10 to divide the test set in order to avoid arbitrary thresholds.
\end{itemize}

\subsubsection{Behavioral and qualitative tests} Our final set of tests is \textit{behavioral} in nature, and tries to capture (with some assumptions) how models differ based on qualitative aspects:

\begin{itemize}
    \item \textbf{Be less wrong}. It is important that RSs maintain a reasonable standard of relevance even when the predictions are not accurate. For instance, if the ground truth for a recommendation is the rap song \textit{‘Humble’} by Kendrick Lamar, a model might suggest another rap song from the same year (\textit{‘The story of O.J.’} by Jay-Z), or a famous pop song from the top chart of that year (\textit{‘Shape of You’} by Ed Sheeran). There is still a substantial difference between these two as the first one is closer to the ground truth than the second. Since this has a great impact on the overall user experience, it is desirable that models test and measure their performance scenarios like the one just described. We use the latent space of tracks to compute the average pairwise cosine distance between the embeddings of the predicted items and the ground truths.
    \item \textbf{Latent diversity}: Diversity is closely tied with the maximization of marginal relevance as a way to acknowledge uncertainty of user intent and to address user utility in terms of discovery \cite{Drosouetal2017}. Diversity is often considered a partial proxy for fairness and it is an important measure of the performance of recommender systems in real world scenarios \cite{KunaverandPozrl2017}. We address diversity using the latent space of tracks testing for model density - where density is defined as the summation of the differences between each point in the prediction space and the mean of the prediction space. Additionally, in order to account also for the ``correctness'' of prediction vectors, we calculate a bias defined as the distance between the ground truth vector and the mean of the prediction vector and weight to penalize for high bias: the final score is computed as $0.3$ * diversity - $0.7$ * bias, where $0.3$ and $0.7$ are weights that we determined empirically to balance diversity and correctness.
\end{itemize}

Please note that since we aim at widening the community contribution to testing, the final code submission for \texttt{EvalRS} includes \textit{as a requirement} that participants contribute at least one custom test, by extending the provided abstraction. 

\subsubsection{Final score}
\label{sec:score}

Since each of the tests above return a score from a potentially unique, non-normal distribution, we need a way to define a \textit{macro-score} for the leaderboard. To define the formula we adopt an empirical approach in two phases:

\begin{enumerate}
    \item \textit{First phase}: scores of individual tests are simply averaged to get the leaderboard macro-score. The purpose of this phase is to gather data on the relative difficulty and utility of the different tests, and get participants comfortable, through harmless iterations, with the dataset and the multi-faceted nature of the challenge.
    \item \textit{Second phase}: after the organizers have evaluated the score distributions for individual tests, they will attach different weights to each test to produce a balanced \textit{macro-score} - i.e. if a test turns out to be easy for most participants, its importance will be counter-biased in the calculation. At the beginning of this phase, participants are asked to update their evaluation script by cloning again the data challenge repository: the purpose for each team becomes now leveraging the insights from the previous phase to optimize their models as much as possible for the leaderboard. \textit{Only scores obtained in this phase are considered for the final prizes}.
\end{enumerate}

\subsection{Methodology}
\label{sec:methdology}

Since the focus of the challenge is a popular public dataset, we implemented a robust evaluation procedure to avoid data leakage and ensure fairness\footnote{To help participants with the implementation, we provide a template script that can be modified with custom model code.}. Our protocol is split in two phases: \textit{local} -- when teams iterate on their solution \textit{during the challenge} - and \textit{remote} -- when organizers verify the submissions at the end and proclaim the winners:

\begin{itemize}
    \item \textit{Local evaluation protocol}: For each fold, the provided script first samples 25\% of the users in the dataset. It then partitions the dataset into training and testing sets using the leave-one-out protocol: the testing set comprises a list of unique users, where the target song for each of them has been picked randomly from their history. The training set is the listening history for these sampled users with their test song removed. Participants' models will be trained and tuned based on their custom logic on the training set, and then evaluated over the test suite (Section \ref{sec:eval}) to provide a final score for each run (Section~\ref{sec:score}); partitioning, training, testing, scoring will be done for a total of $4$ repetitions: the average of the runs will constitute the leaderboard score.
    \item \textit{Remote evaluation protocol}: the organizers will run the code submitted by participants, and repeat the random evaluation loop. The scores thus obtained on the \texttt{EvalRS} test suite will be compared with participants submissions as a sanity check (statistical comparison of means and 95\% bootstrapped CI).
\end{itemize}

Thanks to the provided APIs, participants will be able to run the full evaluation loop locally, as well as update their leaderboard score automatically through the provided script. To ensure a fair and reproducible remote evaluation, final submission should contain a docker image that runs the local evaluation script and produces the desired output \textit{within the maximum allotted time on the target cloud machine}. Please check \texttt{EvalRS} repository for the exact final requirements and up-to-date instructions.

\section{Organization, Community, Impact}

\subsection{Structure and timeline}

\texttt{EvalRS} unfolds in three main phases:

\begin{enumerate}
    \item \textbf{CHALLENGE}: An open challenge phase, where participating teams register for the challenge and work on improving the scores on both standard and behavioral metrics across the two phases explained above (\ref{sec:score}).
    \item \textbf{CFP}: A call for papers, where teams submit a written contribution, describing their system, custom testing, data insights.
    \item \textbf{CONFERENCE}: At the conference, winners will be announced and special prizes for novel testings and oustanding student work will be awarded. During the workshop, we plan to discuss solicited papers and host a round-table with experts on RSs evaluation.
\end{enumerate}

Our \textit{CFP} takes a ``design paper'' perspective, where teams are invited to discuss both how they adapted their initial model to take into account the test suite, and how the tests strengthened their understanding of the target dataset and use case\footnote{As customary in these events, we will involve a small committee from top-tier practitioners and scholars to ensure the quality of the final submissions.}. 

We emphasize the \textit{CFP} and \textit{CONFERENCE} steps as moments to share with the community \textit{additional} tests, error analysis and data insights inspired by \texttt{EvalRS}. By leveraging \textit{RecList}, we not only enable teams to quickly iterate starting from our ideas, but we promise to immediately circulate in the community their testing contribution through a popular open source package. Finally, we plan on using CEUR-WS to publish the accepted papers, as well as drafting a final public report as an additional, actionable artifacts from the challenge.

\subsection{Organizers}

\paragraph{Jacopo Tagliabue} Jacopo Tagliabue was co-founder of Tooso, an Information Retrieval company acquired by Coveo in 2019. As Director of AI at Coveo, he divides his time between product, research, and evangelization: he is Adj. Professor of MLSys at NYU, publishes regularly in top-tier conferences (including NAACL, ACL, RecSys, SIGIR), and is co-organizer of SIGIR eCom. Jacopo was the lead organizer of the SIGIR Data Challenge 2021, spearheading the release of the largest session-based dataset for eCommerce research.

\paragraph{Federico Bianchi}
Federico Bianchi is a postdoctoral researcher at Stanford University. He obtained his Ph.D. in Computer Science at the University of Milano-Bicocca in 2020. His research, ranging from Natural Language Processing methods for textual analytics to recommender systems for the e-commerce has been accepted to major NLP and AI conferences (EACL, NAACL, EMNLP, ACL, AAAI, RecSys) and journals (Cognitive Science, Applied Intelligence, Semantic Web Journals). He co-organized the SIGIR Data Challenge 2021. He frequently releases his research as open-source tools that have collected almost a thousand GitHub stars and been downloaded over 100 thousand times.

\paragraph{Tobias Schnabel}
Tobias Schnabel is a senior researcher in the Productivity+Intelligence group at Microsoft Research. He is interested in improving human-facing machine learning systems in an integrated way, considering not only algorithmic but also human factors. To this end, his research draws from causal inference, reinforcement learning, machine learning, HCI, and decision-making under uncertainty. He was a co-organizer for a WSDM workshop this year and has served as (senior) PC member for a wide array of AI and data science conference (ICML, NeurIPS, WSDM, KDD). Before joining Microsoft, he obtained Ph.D. from the Computer Science Department at Cornell University under Thorsten Joachims.

\paragraph{Giuseppe Attanasio}
Giuseppe Attanasio is a postdoctoral researcher at Bocconi, where he works on large-scale neural architectures for Natural Language Processing. His research focuses on understanding and regularizing models for debiasing and fairness purposes. His research on the topic has been accepted to major NLP conferences (ACL). While working at Bocconi, he is concluding his Ph.D. at the Department of Control and Computer Engineering at Politecnico di Torino.

\paragraph{Ciro Greco}
Ciro Greco was the co-founder and CEO of Tooso, a San Francisco based startup specialized in Information Retrieval. Tooso was acquired in 2019 by Coveo, where he now works as VP or Artificial Intelligence.He holds a Ph.D. in Linguistics and Cognitive Neuroscience at Milano-Bicocca. He worked as visiting scholar at MIT and as a post-doctoral fellow at Ghent University. He published extensively in top-tier conferences (including NAACL, ACL, RecSys, SIGIR) and scientific journals (The Linguistic Review, Cognitive Science, Nature Communications). He was also co-organizer of the SIGIR Data Challenge 2021.

\paragraph{Gabriel de Souza P. Moreira}
Gabriel Moreira is a Sr. Applied Research Scientist at NVIDIA, leading the research efforts of Merlin research team. He had his PhD degree from ITA university, Brazil, with a focus on Deep Learning for RecSys and Session-based recommendation. Before joining NVIDIA, he was lead Data Scientist at CI\&T for 5 years, after working as software engineer for more than a decade. In 2019, he was recognized as a Google Developer Expert (GDE) for Machine Learning. He was part of the NVIDIA teams that won recent RecSys competitions: ACM RecSys Challenge 2020, WSDM WebTour Workshop Challenge 2021 by Booking.com and the SIGIR eCommerce Workshop Data Challenge 2021 by Coveo.

\paragraph{Patrick John Chia}
Patrick John Chia is an Applied Scientist at Coveo. Prior to this, he completed his Master's degree at Imperial College London and spent a year at Massachusetts Institute of Technology (MIT). He was co-organizer of the 2021 SIGIR Data Challenge and has been a speaker on topics at the intersection of Machine Learning and eCommerce (SIGIR eCom, ECNLP at ACL). His latest interests lie in developing AI that has the ability to learn like infants and applying it to creating solutions at Coveo.

\section{Similar Events and Broader Outlook}

The CIKM-related community has shown great interest in themes at the intersection of aligning machine learning with human judgment, rigorous evaluation settings, and fairness, as witnessed by popular Data Challenges and important workshops in top-tier venues. Among recent challenges, the 2021 SIGIR-Ecom Data Challenge, the 2021 Booking Data Challenge, and the 2020 RecSys Challenge are all events centered around the evaluation of RSs, yet still substantially different: for example, the SIGIR Challenge focused on MRR as a success metric \cite{CoveoSIGIR2021}, while the Booking Challenge \cite{DOI_CODE} used top-k accuracy. 

Moreover, the growing interest for rounded evaluation led to the creation of many interesting workshops in recent years, such as \textit{IntRS: Joint Workshop on Interfaces and Human Decision Making for Recommender Systems}, \textit{ImpactRS: Workshop on the Impact of Recommender Systems} and \textit{FAccTRec: Workshop on Responsible Recommendation}. For this reason, we expect \textit{this} challenge to attract a diverse set of practitioners: first, researchers interested in the evaluation of RSs and fairness; second, researchers who proposed a new model and desire to test its generalization abilities on new metrics; third, industrial practitioners that started using \texttt{RecList} after its release in recent months, and already signaled strong support for behavioral testing in their real-world use cases.

\texttt{EvalRS} makes a novel and significant contribution to the community: first, we ask practitioners to ``live and breath'' the problem of evaluation, operationalizing principles and insights through sharable code; second, we embrace a ``build in the open'' approach, as all artifacts from the event will be available to the community as a permanent contribution, in the form of open source code, design papers, and public documentation -- through prizes assigned based on scores, but also outstanding testing and paper contributions, and special awards for students, we hope to actively encourage more practitioners to join the evaluation debate and get a more diverse set of perspectives for our workshop. 

As argued throughout \textit{this} paper, when comparing \texttt{EvalRS} methodology to typical data challenges, we can summarize three important differentiating factors: \textit{first}, we fight public leaderboard overfitting through our randomized evaluation loop; \textit{second}, we discourage complex solutions that cannot be practically used, as our open source code competition provides a fixed (and reasonable) compute budget; \textit{third} and most importantly, with a thorough evaluation with per-group and behavioral tests, we encourage participants to seek non-standard performance and discuss fairness implications. 

We strongly believe these points will lay down the foundation for a first-of-its-kind automatic, shared, identifiable evaluation standard for RSs.

\section{ACKNOWLEDGEMENTS}

\texttt{RecList} is an open source library whose development is supported by forward looking companies in the machine learning community: the organizers wish to thank \textit{Comet}, \textit{Neptune}, \textit{Gantry} for their generous support.\footnote{Please check the project website for more details: \url{https://reclist.io/}.}


\bibliographystyle{ACM-Reference-Format}
\bibliography{sample-base}


\begin{thebibliography}{38}


\ifx \showCODEN    \undefined \def \showCODEN     #1{\unskip}     \fi
\ifx \showDOI      \undefined \def \showDOI       #1{#1}\fi
\ifx \showISBNx    \undefined \def \showISBNx     #1{\unskip}     \fi
\ifx \showISBNxiii \undefined \def \showISBNxiii  #1{\unskip}     \fi
\ifx \showISSN     \undefined \def \showISSN      #1{\unskip}     \fi
\ifx \showLCCN     \undefined \def \showLCCN      #1{\unskip}     \fi
\ifx \shownote     \undefined \def \shownote      #1{#1}          \fi
\ifx \showarticletitle \undefined \def \showarticletitle #1{#1}   \fi
\ifx \showURL      \undefined \def \showURL       {\relax}        \fi
\providecommand\bibfield[2]{#2}
\providecommand\bibinfo[2]{#2}
\providecommand\natexlab[1]{#1}
\providecommand\showeprint[2][]{arXiv:#2}

\bibitem[Kun(2017)]%
        {KunaverandPozrl2017}
 \bibinfo{year}{2017}\natexlab{}.
\newblock \showarticletitle{Diversity in recommender systems – A survey}.
\newblock \bibinfo{journal}{\emph{Knowledge-Based Systems}}
  (\bibinfo{year}{2017}), \bibinfo{pages}{154--162}.
\newblock


\bibitem[Ariu et~al\mbox{.}(2020)]%
        {Ariu2020RegretIO}
\bibfield{author}{\bibinfo{person}{Kaito Ariu}, \bibinfo{person}{Narae Ryu},
  \bibinfo{person}{Seyoung Yun}, {and} \bibinfo{person}{Alexandre
  Prouti{\`e}re}.} \bibinfo{year}{2020}\natexlab{}.
\newblock \showarticletitle{Regret in Online Recommendation Systems}.
\newblock \bibinfo{journal}{\emph{ArXiv}}  \bibinfo{volume}{abs/2010.12363}
  (\bibinfo{year}{2020}).
\newblock


\bibitem[Baigorria~Alonso(2021)]%
        {DOI_CODE}
\bibfield{author}{\bibinfo{person}{Martín Baigorria~Alonso}.}
  \bibinfo{year}{2021}\natexlab{}.
\newblock \showarticletitle{Data Augmentation Using Many-To-Many RNNs for
  Session-Aware Recommender Systems}.
  \bibinfo{howpublished}{\url{https://mbaigorria.github.io/booking-challenge-2021-recsys}}.
  In \bibinfo{booktitle}{\emph{ACM WSDM Workshop on Web Tourism (WSDM
  WebTour’21)}}.
\newblock


\bibitem[Batagelj and Zaveršnik(2011)]%
        {batagelj2002generalized}
\bibfield{author}{\bibinfo{person}{V. Batagelj} {and} \bibinfo{person}{M.
  Zaveršnik}.} \bibinfo{year}{2011}\natexlab{}.
\newblock \showarticletitle{Generalized Cores}.
\newblock \bibinfo{journal}{\emph{Advances in Data Analysis and
  Classification}} \bibinfo{volume}{5}, \bibinfo{number}{2}
  (\bibinfo{year}{2011}), \bibinfo{pages}{129--145}.
\newblock


\bibitem[Bello and Garcia(2021)]%
        {BelloandGarcia2021}
\bibfield{author}{\bibinfo{person}{Pablo Bello} {and} \bibinfo{person}{David
  Garcia}.} \bibinfo{year}{2021}\natexlab{}.
\newblock \showarticletitle{Cultural Divergence in popular music: the
  increasing diversity of music consumption on Spotify across countries}.
\newblock \bibinfo{journal}{\emph{Humanities and Social Sciences
  Communications}} \bibinfo{volume}{8}, \bibinfo{number}{182}
  (\bibinfo{year}{2021}).
\newblock


\bibitem[Bianchi et~al\mbox{.}(2021)]%
        {bianchi-etal-2021-query2prod2vec}
\bibfield{author}{\bibinfo{person}{Federico Bianchi}, \bibinfo{person}{Jacopo
  Tagliabue}, {and} \bibinfo{person}{Bingqing Yu}.}
  \bibinfo{year}{2021}\natexlab{}.
\newblock \showarticletitle{{Q}uery2{P}rod2{V}ec: Grounded Word Embeddings for
  e{C}ommerce}. In \bibinfo{booktitle}{\emph{Proceedings of the 2021 Conference
  of the North American Chapter of the Association for Computational
  Linguistics: Human Language Technologies: Industry Papers}}.
  \bibinfo{publisher}{Association for Computational Linguistics},
  \bibinfo{address}{Online}, \bibinfo{pages}{154--162}.
\newblock
\urldef\tempurl%
\url{https://doi.org/10.18653/v1/2021.naacl-industry.20}
\showDOI{\tempurl}


\bibitem[Briand et~al\mbox{.}(2020)]%
        {2106.03819}
\bibfield{author}{\bibinfo{person}{Léa Briand}, \bibinfo{person}{Guillaume
  Salha-Galvan}, \bibinfo{person}{Walid Bendada}, \bibinfo{person}{Mathieu
  Morlon}, {and} \bibinfo{person}{Viet-Anh Tran}.}
  \bibinfo{year}{2020}\natexlab{}.
\newblock \showarticletitle{A Semi-Personalized System for User Cold Start
  Recommendation on Music Streaming Apps}.
\newblock
\urldef\tempurl%
\url{arXiv:2106.03819}
\showURL{%
\tempurl}


\bibitem[Castillo(2019)]%
        {Castillo2019}
\bibfield{author}{\bibinfo{person}{Carlos Castillo}.}
  \bibinfo{year}{2019}\natexlab{}.
\newblock \showarticletitle{Fairness and Transparency in Ranking}. In
  \bibinfo{booktitle}{\emph{ACM SIGIR ForumVolume}},
  Vol.~\bibinfo{volume}{Volume 52}. \bibinfo{pages}{64 -- 71}.
\newblock
Issue 2.
\urldef\tempurl%
\url{https://doi.org/10.1145/3308774.3308783}
\showURL{%
\tempurl}


\bibitem[Chia et~al\mbox{.}(2021)]%
        {DBLP:journals/corr/abs-2111-09963}
\bibfield{author}{\bibinfo{person}{Patrick~John Chia}, \bibinfo{person}{Jacopo
  Tagliabue}, \bibinfo{person}{Federico Bianchi}, \bibinfo{person}{Chloe He},
  {and} \bibinfo{person}{Brian Ko}.} \bibinfo{year}{2021}\natexlab{}.
\newblock \showarticletitle{Beyond {NDCG:} behavioral testing of recommender
  systems with RecList}.
\newblock \bibinfo{journal}{\emph{CoRR}}  \bibinfo{volume}{abs/2111.09963}
  (\bibinfo{year}{2021}).
\newblock
\showeprint[arXiv]{2111.09963}
\urldef\tempurl%
\url{https://arxiv.org/abs/2111.09963}
\showURL{%
\tempurl}


\bibitem[Dacrema et~al\mbox{.}(2019)]%
        {10.1145/3298689.3347058}
\bibfield{author}{\bibinfo{person}{Maurizio~Ferrari Dacrema},
  \bibinfo{person}{Paolo Cremonesi}, {and} \bibinfo{person}{Dietmar Jannach}.}
  \bibinfo{year}{2019}\natexlab{}.
\newblock \showarticletitle{Are We Really Making Much Progress? A Worrying
  Analysis of Recent Neural Recommendation Approaches}. In
  \bibinfo{booktitle}{\emph{Proceedings of the 13th ACM Conference on
  Recommender Systems}} (Copenhagen, Denmark) \emph{(\bibinfo{series}{RecSys
  '19})}. \bibinfo{publisher}{Association for Computing Machinery},
  \bibinfo{address}{New York, NY, USA}, \bibinfo{pages}{101–109}.
\newblock
\showISBNx{9781450362436}
\urldef\tempurl%
\url{https://doi.org/10.1145/3298689.3347058}
\showDOI{\tempurl}


\bibitem[Drosou et~al\mbox{.}(2017)]%
        {Drosouetal2017}
\bibfield{author}{\bibinfo{person}{Marina Drosou}, \bibinfo{person}{H.V.
  Jagadish}, \bibinfo{person}{Evaggelia Pitoura}, {and} \bibinfo{person}{Julia
  Stoyanovich}.} \bibinfo{year}{2017}\natexlab{}.
\newblock \showarticletitle{Diversity in big data: A review}.
\newblock \bibinfo{journal}{\emph{Big data 5.2}} (\bibinfo{year}{2017}),
  \bibinfo{pages}{73--84}.
\newblock


\bibitem[Flexer et~al\mbox{.}(2012)]%
        {Flexeretal2012}
\bibfield{author}{\bibinfo{person}{Arthur Flexer}, \bibinfo{person}{Dominik
  Schnitzer}, {and} \bibinfo{person}{Jan Schlueter}.}
  \bibinfo{year}{2012}\natexlab{}.
\newblock \showarticletitle{A MIREX Meta-analysis of Hubness in Audio Music
  Similarity}.
\newblock


\bibitem[Harper and Konstan(2015)]%
        {10.1145/2827872}
\bibfield{author}{\bibinfo{person}{F.~Maxwell Harper} {and}
  \bibinfo{person}{Joseph~A. Konstan}.} \bibinfo{year}{2015}\natexlab{}.
\newblock \showarticletitle{The MovieLens Datasets: History and Context}.
\newblock \bibinfo{journal}{\emph{ACM Trans. Interact. Intell. Syst.}}
  \bibinfo{volume}{5}, \bibinfo{number}{4}, Article \bibinfo{articleno}{19}
  (\bibinfo{date}{Dec.} \bibinfo{year}{2015}), \bibinfo{numpages}{19}~pages.
\newblock
\showISSN{2160-6455}
\urldef\tempurl%
\url{https://doi.org/10.1145/2827872}
\showDOI{\tempurl}


\bibitem[Hendriksen et~al\mbox{.}(2020)]%
        {Hendriksenetal2020}
\bibfield{author}{\bibinfo{person}{Mariya Hendriksen}, \bibinfo{person}{Ernst
  Kuiper}, \bibinfo{person}{Pim Nauts}, \bibinfo{person}{Sebastian Schelter},
  {and} \bibinfo{person}{Maarten de Rijke}.} \bibinfo{year}{2020}\natexlab{}.
\newblock \showarticletitle{Analyzing and Predicting Purchase Intent in
  E-commerce: Anonymous vs. Identified Customers}.
\newblock
\urldef\tempurl%
\url{https://arxiv.org/abs/2012.08777}
\showURL{%
\tempurl}


\bibitem[Jannach and Ludewig(2017)]%
        {jannach2017recurrent}
\bibfield{author}{\bibinfo{person}{Dietmar Jannach} {and}
  \bibinfo{person}{Malte Ludewig}.} \bibinfo{year}{2017}\natexlab{}.
\newblock \showarticletitle{When recurrent neural networks meet the
  neighborhood for session-based recommendation}. In
  \bibinfo{booktitle}{\emph{Proceedings of the Eleventh ACM Conference on
  Recommender Systems}}. \bibinfo{pages}{306--310}.
\newblock


\bibitem[Ke~Yang(2017)]%
        {YangandStoyanovich2016}
\bibfield{author}{\bibinfo{person}{Julia~Stoyanovich Ke~Yang}.}
  \bibinfo{year}{2017}\natexlab{}.
\newblock \showarticletitle{Measuring Fairness in Ranked Outputs}. In
  \bibinfo{booktitle}{\emph{SSDBM 2017: Proceedings of the 29th International
  Conference on Scientific and Statistical Database Management}}.
  \bibinfo{pages}{1 -- 6}.
\newblock
\urldef\tempurl%
\url{https://doi.org/10.1145/3085504.3085526}
\showURL{%
\tempurl}


\bibitem[Kotkov et~al\mbox{.}(2016)]%
        {Kotkov2016ChallengesOS}
\bibfield{author}{\bibinfo{person}{Denis Kotkov}, \bibinfo{person}{Jari
  Veijalainen}, {and} \bibinfo{person}{Shuaiqiang Wang}.}
  \bibinfo{year}{2016}\natexlab{}.
\newblock \showarticletitle{Challenges of Serendipity in Recommender Systems}.
  In \bibinfo{booktitle}{\emph{WEBIST}}.
\newblock


\bibitem[Kouki et~al\mbox{.}(2020)]%
        {10.1145/3383313.3412235}
\bibfield{author}{\bibinfo{person}{Pigi Kouki}, \bibinfo{person}{Ilias
  Fountalis}, \bibinfo{person}{Nikolaos Vasiloglou}, \bibinfo{person}{Xiquan
  Cui}, \bibinfo{person}{Edo Liberty}, {and} \bibinfo{person}{Khalifeh
  Al~Jadda}.} \bibinfo{year}{2020}\natexlab{}.
\newblock \showarticletitle{From the Lab to Production: A Case Study of
  Session-Based Recommendations in the Home-Improvement Domain}. In
  \bibinfo{booktitle}{\emph{Fourteenth ACM Conference on Recommender Systems}}
  (Virtual Event, Brazil) \emph{(\bibinfo{series}{RecSys '20})}.
  \bibinfo{publisher}{Association for Computing Machinery},
  \bibinfo{address}{New York, NY, USA}, \bibinfo{pages}{140–149}.
\newblock
\showISBNx{9781450375832}
\urldef\tempurl%
\url{https://doi.org/10.1145/3383313.3412235}
\showDOI{\tempurl}


\bibitem[Kowald et~al\mbox{.}(2020)]%
        {Kowaldetal2020}
\bibfield{author}{\bibinfo{person}{Dominik Kowald}, \bibinfo{person}{Markus
  Schedl}, {and} \bibinfo{person}{Elisabeth Lex}.}
  \bibinfo{year}{2020}\natexlab{}.
\newblock \showarticletitle{The Unfairness of Popularity Bias in Music
  Recommendation: A Reproducibility Study}.
\newblock \bibinfo{journal}{\emph{European conference on information
  retrieval}} (\bibinfo{year}{2020}).
\newblock


\bibitem[{Krista Garcia}(2018)]%
        {emarketer}
\bibfield{author}{\bibinfo{person}{{Krista Garcia}}.}
  \bibinfo{year}{2018}\natexlab{}.
\newblock \bibinfo{booktitle}{\emph{The Impact of Product Recommendations}}.
\newblock
\urldef\tempurl%
\url{https://www.emarketer.com/content/the-impact-of-product-recommendations}
\showURL{%
Retrieved November 9, 2021 from \tempurl}


\bibitem[Ludewig and Jannach(2018)]%
        {ludewig2018evaluation}
\bibfield{author}{\bibinfo{person}{Malte Ludewig} {and}
  \bibinfo{person}{Dietmar Jannach}.} \bibinfo{year}{2018}\natexlab{}.
\newblock \showarticletitle{Evaluation of session-based recommendation
  algorithms}.
\newblock \bibinfo{journal}{\emph{User Modeling and User-Adapted Interaction}}
  \bibinfo{volume}{28}, \bibinfo{number}{4-5} (\bibinfo{year}{2018}),
  \bibinfo{pages}{331--390}.
\newblock


\bibitem[Moins et~al\mbox{.}(2020)]%
        {10.1145/3383313.3412263}
\bibfield{author}{\bibinfo{person}{Th\'{e}o Moins}, \bibinfo{person}{Daniel
  Aloise}, {and} \bibinfo{person}{Simon~J. Blanchard}.}
  \bibinfo{year}{2020}\natexlab{}.
\newblock \showarticletitle{RecSeats: A Hybrid Convolutional Neural Network
  Choice Model for Seat Recommendations at Reserved Seating Venues}. In
  \bibinfo{booktitle}{\emph{Fourteenth ACM Conference on Recommender Systems}}
  (Virtual Event, Brazil) \emph{(\bibinfo{series}{RecSys '20})}.
  \bibinfo{publisher}{Association for Computing Machinery},
  \bibinfo{address}{New York, NY, USA}, \bibinfo{pages}{309–317}.
\newblock
\showISBNx{9781450375832}
\urldef\tempurl%
\url{https://doi.org/10.1145/3383313.3412263}
\showDOI{\tempurl}


\bibitem[Moreira et~al\mbox{.}(2021)]%
        {transformers4rec2021}
\bibfield{author}{\bibinfo{person}{Gabriel de Souza~Pereira Moreira},
  \bibinfo{person}{Sara Rabhi}, \bibinfo{person}{Jeong~Min Lee},
  \bibinfo{person}{Ronay Ak}, {and} \bibinfo{person}{Even Oldridge}.}
  \bibinfo{year}{2021}\natexlab{}.
\newblock \showarticletitle{Transformers4Rec: Bridging the Gap between NLP and
  Sequential/Session-Based Recommendation}. In
  \bibinfo{booktitle}{\emph{Fifteenth ACM Conference on Recommender Systems}}.
  \bibinfo{pages}{143--153}.
\newblock


\bibitem[O'Mahony et~al\mbox{.}(2004)]%
        {OMahony2018}
\bibfield{author}{\bibinfo{person}{Michael O'Mahony}, \bibinfo{person}{Neil
  Hurley}, \bibinfo{person}{Nicholas Kushmerick}, {and}
  \bibinfo{person}{Guénolé Silvestre}.} \bibinfo{year}{2004}\natexlab{}.
\newblock \showarticletitle{Collaborative recommendation: A robustness
  analysis}, Vol.~\bibinfo{volume}{4}.
\newblock
Issue 4.
\urldef\tempurl%
\url{https://doi.org/10.1145/1031114.1031116}
\showURL{%
\tempurl}


\bibitem[Rashed et~al\mbox{.}(2020)]%
        {Rashed2020MultiRecAM}
\bibfield{author}{\bibinfo{person}{Ahmed Rashed}, \bibinfo{person}{Shayan
  Jawed}, \bibinfo{person}{Lars Schmidt-Thieme}, {and} \bibinfo{person}{Andre
  Hintsches}.} \bibinfo{year}{2020}\natexlab{}.
\newblock \showarticletitle{MultiRec: A Multi-Relational Approach for Unique
  Item Recommendation in Auction Systems}.
\newblock \bibinfo{journal}{\emph{Fourteenth ACM Conference on Recommender
  Systems}} (\bibinfo{year}{2020}).
\newblock


\bibitem[Ribeiro et~al\mbox{.}(2020)]%
        {Ribeiro2020BeyondAB}
\bibfield{author}{\bibinfo{person}{Marco~Tulio Ribeiro},
  \bibinfo{person}{Tongshuang~Sherry Wu}, \bibinfo{person}{Carlos Guestrin},
  {and} \bibinfo{person}{Sameer Singh}.} \bibinfo{year}{2020}\natexlab{}.
\newblock \showarticletitle{Beyond Accuracy: Behavioral Testing of NLP Models
  with CheckList}. In \bibinfo{booktitle}{\emph{ACL}}.
\newblock


\bibitem[Saxena and Jain(2021)]%
        {SaxenaandJein2020}
\bibfield{author}{\bibinfo{person}{Shrikant Saxena} {and}
  \bibinfo{person}{Shweta Jain}.} \bibinfo{year}{2021}\natexlab{}.
\newblock \showarticletitle{Exploring and Mitigating Gender Bias in Recommender
  Systems with Explicit Feedback}.
\newblock  (\bibinfo{year}{2021}).
\newblock
\urldef\tempurl%
\url{arXiv preprint arXiv:2112.02530}
\showURL{%
\tempurl}


\bibitem[Schedl(2016)]%
        {10.1145/2911996.2912004}
\bibfield{author}{\bibinfo{person}{Markus Schedl}.}
  \bibinfo{year}{2016}\natexlab{}.
\newblock \showarticletitle{The LFM-1b Dataset for Music Retrieval and
  Recommendation}. In \bibinfo{booktitle}{\emph{Proceedings of the 2016 ACM on
  International Conference on Multimedia Retrieval}} (New York, New York, USA)
  \emph{(\bibinfo{series}{ICMR '16})}. \bibinfo{publisher}{Association for
  Computing Machinery}, \bibinfo{address}{New York, NY, USA},
  \bibinfo{pages}{103–110}.
\newblock
\showISBNx{9781450343596}
\urldef\tempurl%
\url{https://doi.org/10.1145/2911996.2912004}
\showDOI{\tempurl}


\bibitem[Schedl(2017)]%
        {Schedl2017InvestigatingCM}
\bibfield{author}{\bibinfo{person}{Markus Schedl}.}
  \bibinfo{year}{2017}\natexlab{}.
\newblock \showarticletitle{Investigating country-specific music preferences
  and music recommendation algorithms with the LFM-1b dataset}.
\newblock \bibinfo{journal}{\emph{International Journal of Multimedia
  Information Retrieval}}  \bibinfo{volume}{6} (\bibinfo{year}{2017}),
  \bibinfo{pages}{71 -- 84}.
\newblock


\bibitem[Sun et~al\mbox{.}(2020)]%
        {sun2020we}
\bibfield{author}{\bibinfo{person}{Zhu Sun}, \bibinfo{person}{Di Yu},
  \bibinfo{person}{Hui Fang}, \bibinfo{person}{Jie Yang},
  \bibinfo{person}{Xinghua Qu}, \bibinfo{person}{Jie Zhang}, {and}
  \bibinfo{person}{Cong Geng}.} \bibinfo{year}{2020}\natexlab{}.
\newblock \showarticletitle{Are we evaluating rigorously? benchmarking
  recommendation for reproducible evaluation and fair comparison}. In
  \bibinfo{booktitle}{\emph{Fourteenth ACM conference on recommender systems}}.
  \bibinfo{pages}{23--32}.
\newblock


\bibitem[Tagliabue(2021)]%
        {10.1145/3460231.3474604}
\bibfield{author}{\bibinfo{person}{Jacopo Tagliabue}.}
  \bibinfo{year}{2021}\natexlab{}.
\newblock \bibinfo{booktitle}{\emph{You Do Not Need a Bigger Boat:
  Recommendations at Reasonable Scale in a (Mostly) Serverless and Open
  Stack}}.
\newblock \bibinfo{publisher}{Association for Computing Machinery},
  \bibinfo{address}{New York, NY, USA}, \bibinfo{pages}{598–600}.
\newblock
\showISBNx{9781450384582}
\urldef\tempurl%
\url{https://doi.org/10.1145/3460231.3474604}
\showURL{%
\tempurl}


\bibitem[Tagliabue et~al\mbox{.}(2021)]%
        {CoveoSIGIR2021}
\bibfield{author}{\bibinfo{person}{Jacopo Tagliabue}, \bibinfo{person}{Ciro
  Greco}, \bibinfo{person}{Jean-Francis Roy}, \bibinfo{person}{Federico
  Bianchi}, \bibinfo{person}{Giovanni Cassani}, \bibinfo{person}{Bingqing Yu},
  {and} \bibinfo{person}{Patrick~John Chia}.} \bibinfo{year}{2021}\natexlab{}.
\newblock \showarticletitle{SIGIR 2021 E-Commerce Workshop Data Challenge}. In
  \bibinfo{booktitle}{\emph{SIGIR eCom 2021}}.
\newblock


\bibitem[Tagliabue et~al\mbox{.}(2020)]%
        {10.1145/3383313.3411477}
\bibfield{author}{\bibinfo{person}{Jacopo Tagliabue}, \bibinfo{person}{Bingqing
  Yu}, {and} \bibinfo{person}{Federico Bianchi}.}
  \bibinfo{year}{2020}\natexlab{}.
\newblock \bibinfo{booktitle}{\emph{The Embeddings That Came in From the Cold:
  Improving Vectors for New and Rare Products with Content-Based Inference}}.
\newblock \bibinfo{publisher}{Association for Computing Machinery},
  \bibinfo{address}{New York, NY, USA}, \bibinfo{pages}{577–578}.
\newblock
\showISBNx{9781450375832}
\urldef\tempurl%
\url{https://doi.org/10.1145/3383313.3411477}
\showURL{%
\tempurl}


\bibitem[Twohey and Dance(2022)]%
        {NYTIMES_AMAZON}
\bibfield{author}{\bibinfo{person}{Megan Twohey} {and}
  \bibinfo{person}{Gabriel~J.X. Dance}.} \bibinfo{year}{2022}\natexlab{}.
\newblock \bibinfo{booktitle}{\emph{Lawmakers Press Amazon on Sales of Chemical
  Used in Suicides}}.
\newblock
\urldef\tempurl%
\url{https://www.nytimes.com/2022/02/04/technology/amazon-suicide-poison-preservative.html}
\showURL{%
\tempurl}


\bibitem[Wang et~al\mbox{.}(2019)]%
        {wang2019neural}
\bibfield{author}{\bibinfo{person}{Xiang Wang}, \bibinfo{person}{Xiangnan He},
  \bibinfo{person}{Meng Wang}, \bibinfo{person}{Fuli Feng}, {and}
  \bibinfo{person}{Tat-Seng Chua}.} \bibinfo{year}{2019}\natexlab{}.
\newblock \showarticletitle{Neural graph collaborative filtering}. In
  \bibinfo{booktitle}{\emph{Proceedings of the 42nd international ACM SIGIR
  conference on Research and development in Information Retrieval}}.
  \bibinfo{pages}{165--174}.
\newblock


\bibitem[Zamani et~al\mbox{.}(2019)]%
        {10.1145/3344257}
\bibfield{author}{\bibinfo{person}{Hamed Zamani}, \bibinfo{person}{Markus
  Schedl}, \bibinfo{person}{Paul Lamere}, {and} \bibinfo{person}{Ching-Wei
  Chen}.} \bibinfo{year}{2019}\natexlab{}.
\newblock \showarticletitle{An Analysis of Approaches Taken in the ACM RecSys
  Challenge 2018 for Automatic Music Playlist Continuation}.
\newblock \bibinfo{journal}{\emph{ACM Trans. Intell. Syst. Technol.}}
  \bibinfo{volume}{10}, \bibinfo{number}{5}, Article \bibinfo{articleno}{57}
  (\bibinfo{date}{Sept.} \bibinfo{year}{2019}), \bibinfo{numpages}{21}~pages.
\newblock
\showISSN{2157-6904}
\urldef\tempurl%
\url{https://doi.org/10.1145/3344257}
\showDOI{\tempurl}


\bibitem[Zehlike et~al\mbox{.}(2020)]%
        {Zehlikeetal2021}
\bibfield{author}{\bibinfo{person}{Meike Zehlike}, \bibinfo{person}{Ke Yang},
  {and} \bibinfo{person}{Julia Stoyanovich}.} \bibinfo{year}{2020}\natexlab{}.
\newblock \showarticletitle{Fairness in Ranking: A Survey}. In
  \bibinfo{booktitle}{\emph{TBD. ACM}}. \bibinfo{pages}{1--58}.
\newblock
\urldef\tempurl%
\url{https://arxiv.org/pdf/2103.14000.pdf}
\showURL{%
\tempurl}


\bibitem[Òscar Celma and Cano(2008)]%
        {CelmaandCano2008}
\bibfield{author}{\bibinfo{person}{Òscar Celma} {and} \bibinfo{person}{Pedro
  Cano}.} \bibinfo{year}{2008}\natexlab{}.
\newblock \showarticletitle{From hits to niches? or how popular artists can
  bias music recommendation and discovery}. In
  \bibinfo{booktitle}{\emph{Proceedings of the 2nd KDD Workshop on Large-Scale
  Recommender Systems and the Netflix Prize Competition}}.
\newblock
\urldef\tempurl%
\url{https://citeseerx.ist.psu.edu/viewdoc/download?doi=10.1.1.168.5009&rep=rep1&type=pdf}
\showURL{%
\tempurl}


\end{thebibliography}

\end{document}